\documentclass[twocolumn]{aastex62}
\shorttitle{Faraday Tomography with Sparse Modeling}
\shortauthors{K.~Akiyama et al.}
\definecolor{MyDarkBlue}{rgb}{0,0.08,0.5}
\definecolor{MyDarkRed}{rgb}{0.7,0.02,0.02}
\definecolor{MyDarkGreen}{rgb}{0.0,0.7,0.0}

\newcommand{\nrao}{National Radio Astronomy Observatory, 520 Edgemont Rd, Charlottesville, VA 22903, USA}

\newcommand{\naojmiz}{Mizusawa VLBI Observatory, National Astronomical Observatory Japan, 2-21-1 Osawa, Mitaka, Tokyo 181-8588, Japan}
\newcommand{\naojcfca}{Center for Computational Astrophysics, National Astronomical Observatory Japan, 2-21-1 Osawa, Mitaka, Tokyo 181-8588, Japan}
\newcommand{\haystack}{Massachusetts Institute of Technology, Haystack Observatory, 99 Millstone Rd, Westford, MA 01886, USA}
\newcommand{\bhi}{Black Hole Initiative, Harvard University, 20 Garden Street, Cambridge, MA 02138, USA}
\newcommand{\ism}{The Institute of Statistical Mathematics, 10-3 Midori-cho, Tachikawa, Tokyo 190-8562, Japan}

\newcommand{\kumamoto}{Faculty of Advanced Science and Technology, Kumamoto University, 2-39-1, Kurokami, Kumamoto 860-8555, Japan}
\newcommand{\kagoshima}{School of Science, Kagoshima University, 1-21-35, Korimoto, Kagoshima, Kagoshima 890-0065, Japan}
\newcommand{\radboud}{Department of Astrophysics/IMAPP, Radboud University Nijmegen, PO Box 9010, NL-6500 GL Nijmegen, the Netherland}

\usepackage{amsmath}

\begin{document}
\title{Faraday Tomography with Sparse Modeling}
%
\correspondingauthor{Kazunori Akiyama}
\email{kakiyama@mit.edu}
\author{Kazunori Akiyama}
\altaffiliation{NRAO Jansky Fellow}
\affil{\nrao}
\affil{\haystack}
\affil{\bhi}
\affil{\naojmiz}
\author{Takuya Akahori}
\affil{\naojmiz}
\author{Yoshimitsu Miyashita}
\altaffiliation{JSPS Fellow}
\affil{\kumamoto}
\author{Shinsuke Ideguchi}
\affil{\naojcfca}
\affil{\radboud}

\author{Ryosuke Yamaguchi}
\affil{\kagoshima}
\author{Shiro Ikeda}
\affil{\ism}
\author{Keitaro Takahashi}
\affil{\kumamoto}

%
%
\begin{abstract}
Faraday tomography (or rotation measure synthesis) is a procedure to convert linear polarization spectra into the Faraday dispersion function, which provides us with unique information of magneto-ionic media along the line of sight. Mathematical formulation of Faraday tomography is similar to polarimetric imaging of radio interferometry, where many new methods have been actively developed and shown to outperform the standard CLEAN approaches. In this paper, we propose a sparse reconstruction technique to Faraday tomography. This technique is being developed for interferometric imaging and utilizes computationally less expensive convex regularization functions such as $\ell_1$-norm and total variation (TV) or total squared variation (TSV). The proposed technique solves a convex optimization, and therefore its solution is determined uniquely regardless of the initial condition for given regularization parameters that can be optimized by data themselves. Using a physically-motivated model of turbulent galactic magnetized plasma, we demonstrate that the proposed technique outperforms RM-CLEAN and provides higher-fidelity reconstruction. The proposed technique would be a powerful tool in broadband polarimetry with the Square Kilometre Array (SKA) and its precursors.
\end{abstract}

\keywords{radio continuum: ISM --- radio continuum: galaxies --- ISM: magnetic fields --- magnetic fields --- polarization --- techniques: interferometric}

\section{Introduction}\label{sec:1}

Linearly-polarized emission including synchrotron radiation from relativistic electrons is a unique tracer of magnetic field in the Universe. It experiences Faraday rotation and imprints physical information of magneto-ionic plasma along the line of sight as the Faraday depth (or alternatively called as the Faraday rotation measure). There were a lot of surveys or individual observations of the Faraday depth in the last several decades, revealing the role of magnetic field in various astronomical objects \citep[see][for a review]{akahori2018a}.

Modern broadband radio polarimetry advances the study of the Faraday depth to a new discovery space called the Faraday dispersion function (FDF), which is a complex linear-polarization spectrum as a function of the Faraday depth. The FDF provides tomographic information of polarized sources and magneto-ionic media along the line of sight. Reconstruction of the FDF is named Faraday tomography or Faraday rotation measure (RM) synthesis \citep{burn1966, gaensler2004, beck2004, brentjens2005} (see Section~\ref{sec:2}). 

Faraday tomography, however, suffers from the quality of reconstruction \citep{andrecut2012, kumazaki2014, sun2015} due primarily to narrow, limited, and unevenly-sampled observational data in wavelength squared \citep[e.g.,][]{beck2012, akahori2014}. Several advanced techniques have been studied to improve the reconstruction, e.g., RM CLEAN \citep{heald2009}, QU-fitting \citep{osullivan2012}, and compressed-sensing (CS)-based technique \citep{li2011b, andrecut2012} (see Section~\ref{sec:3}).

Faraday tomography shares a common mathematical formulation with polarimetric imaging of radio interferometry \citep[][and also Section \ref{sec:2}]{thompson2017}. In the recent several years, new polarimetric imaging techniques have been actively developed for the next generation very long baseline interferometers such as the Event Horizon Telescope \citep[EHT;][]{doeleman2009}, whose frequency-sample coverages are sparse and also whose primary target sources have a complicated structure on scales of their angular resolutions represented by the black hole shadows in the Galactic center Sgr A* \citep[see, e.g.][]{chael2016, kuramochi2018} and nearby radio galaxy M87 \citep[see, e.g.][]{akiyama2017a, akiyama2017b}. These challenges, which would be common for Faraday tomography with the next generation interferometers, provide active developments of many new high-fidelity polarimetric imaging techniques \citep{chael2016, coughlan2016, akiyama2017a, birdi2018}, consistently outperforming the traditional CLEAN approaches and its variants.

In this work, we present an application of a new radio interferometric imaging technique, often called {\it sparse modeling} \citep{honma2014, ikeda2016, akiyama2017a, akiyama2017b, kuramochi2018}, to Faraday tomography. It is a CS-based technique utilizing sparse regularization on the image itself and its gradient and provides a capability of high-quality superresolution and full polarimeteric imaging \citep{akiyama2017b}. This paper is organized as follows. We define basic equations of Faraday tomography and give an overview of previous work in Section \ref{sec:2}. We then present more detailed general mathematical background of the FDF reconstruction and our techniques in Section \ref{sec:3}. An example application of our technique followed by our discussion is presented in Section \ref{sec:4}. Finally, we briefly summarize this work and some future prospects in Section \ref{sec:5}.

\section{Faraday tomography}\label{sec:2}


Polarized emission is described with four Stokes parameters, $I$, $Q$, $U$, and $V$, which are all real numbers. Stokes $I$ represents the total intensity of the emission, while $Q$ and $U$ represent linear polarization and $V$ represents circular polarization. Stokes $Q$ and $U$ are often combined into the complex quantity $P=Q+iU=|P|e^{i2\chi}$, where $|P|$ and $\chi =\arg(P)/2$ are the linear-polarization intensity and the electric-vector polarization angle (EVPA), respectively. The EVPA propagating through magnetized plasma is rotated by
\begin{equation}
\Delta \chi = \phi \lambda ^ 2
\end{equation}
where $\lambda$ is the wavelength of the emission. The factor $\phi$ is the Faraday depth given by \cite[e.g.][]{pacholczyk1970}
\begin{equation}
\phi \approx 811.9\,\, ({\rm rad\,m^{-2}}) \int ^{\infty}_{0}  \left( \frac{n_e}{{\rm cm^{-2}}} \right) \left( \frac{B_{||}}{{\rm \mu G}} \right) \left( \frac{dr}{{\rm kpc}} \right),
\end{equation}
where $n_e$ is the electron density, and $B_{||}$ and $r$ are the magnetic-field strength and physical depth of the magnetized plasma along the line of sight, respectively. Traditionally, $\phi$ has been estimated by measuring $\Delta \chi$ in multiband polarimetry. 

A linear polarization $\tilde{P}$ is Faraday-rotated by $\phi \lambda^2$ in EVPA, when it passes through the plasma with $\phi$. Thus, the observed linear polarization can be described as $\tilde{P}e^{i2\phi \lambda^2}$. Considering a general case that we observe superposition of intrinsic polarizations at multiple Faraday depths, the observed linear-polarization spectrum $P(\lambda^2)$ is given by
\begin{equation}
P(\lambda ^2) = \int^{\infty}_{-\infty} \tilde{P}(\phi)e^{i2\phi \lambda^2} d\phi, \label{eq:obseq}
\end{equation}
This $\tilde{P}(\phi)$, the intrinsic linear polarization at $\phi$, is named the Faraday dispersion function (FDF), which can be obtained from inverse Fourier transformation of $P(\lambda ^2)$ \citep{burn1966, brentjens2005}. 

Equation~(\ref{eq:obseq}) implies that Faraday tomography is an under-determined ill-posed problem, since the sampling and coverage of linear polarization in wavelength squared are both finite and imperfect in real observations. In other words, the infinite number of possible FDFs satisfies the observational equation (Equation \ref{eq:obseq}) even in an ultimate case without any noise. Therefore, Faraday tomography requires a certain prior assumption for the FDF to pick up a reasonable and likely solution. This mathematical description and properties are common with radio interferometric (polarimetric) imaging \citep[e.g.,][]{thompson2017}, which can be described by replacing the observed spectra $P(\lambda ^2)$ by Stokes visibilities data sets and the FDF to be solved by the corresponding Stokes intensity images. 

If the expected FDF is known from prior knowledges, a promising way is to fit the polarization spectrum with some parametrized models \citep[e.g., {\it QU-fitting};][]{osullivan2012}. The fitting can be done by using traditional likelihood approaches including the least square method \citep[e.g.][]{osullivan2012, ideguchi2014a, ozawa2015, kaczmarek2017}, Bayesian approaches like the Marcov Chain Monte Carlo \citep[MCMC;][]{miyashita2017, schnitzeler2018, sakemi2018}, or the Convolution Neural Network \citep{brown2017}.

Otherwise, if the property of the FDF is not clear, model-independent reconstruction is a more generic approach (see Section \ref{sec:3.2}). A popular prior assumption is the sparsity of solution, where the solution consists of a small number of non-zero elements (see Section \ref{sec:3.3}). A classical approach of sparse reconstruction is CLEAN \citep[e.g.][]{hogbom1974}, which is independently rediscovered as the Matching Pursuit algorithm \citep{mallat1993}. CLEAN has been the standard approach in radio interferometric imaging \citep[see][]{thompson2017}, and the CLEAN algorithm has been successfully exported to Faraday tomography as RM-CLEAN \citep[e.g.][]{heald2009, anderson2016, michilli2018}. 

A state-of-the-art sparse reconstruction is based on the compressed-sensing theory \citep[CS; e.g.][]{donoho2006,candes2006}. The CS-based techniques directly solve the observational equation (Equation \ref{eq:obseq}) and use the convex regularization functions which pick up a sparse solution on some basis from infinite number of possible solutions. \citet{wiaux2009a, wiaux2009b} presented their pioneering works of CS-based techniques in radio astronomy, followed by \citet{li2011b} and \citet{andrecut2012} in Faraday tomography. They utilized sparse regularization on the FDF itself or some other dictionary basis, and demonstrated many advantages of CS-based techniques compared to CLEAN. More detailed mathematical background of the CS-based techniques including the proposed technique and their relation to RM CLEAN is described in Section \ref{sec:3}.

\section{Regularized Maximum-likelihood Sparse Reconstruction}\label{sec:3}


\subsection{Observational Equations}\label{sec:3.1}

We start with formulating the observational equation of Faraday tomography. While a linear-polarization spectrum is measured as a discrete function of the wavelength squared, ${\bf \lambda }^2=\{\lambda ^2_j,\,j=1,2,...,M\}$, the FDF can be approximated by a pixelated version as a function of the Faraday depth, ${\bf \phi}=\{\phi _j,\,j=1,2,...,N\}$. Therefore, the observational equation of Faraday tomography is discretized into the following complex linear equation,
\begin{equation}
{\bf P} = {\bf F}\tilde{\bf P}, \label{eq:obseq1}
\end{equation}
where ${\bf P}=\{P _j,\,j=1,2,...,M\}$ is the measured linear-polarization spectrum, which can be expressed as ${\bf P} = {\bf Q} + i {\bf U}=\{Q_j\} + i\{U_j\}$ using the Stokes parameters. The FDF can be described likewise as $\tilde{\bf{P}} = \{\tilde{P}_j\}=\{\tilde{Q}_j\} + i\{\tilde{U}_j\}$. Fourier Matrix ${\bf F}$ is given by ${\bf F} = \{F_{jk}\} = \{\exp(i2\phi_j\lambda _k ^2)\}$. 

The observational equation can be reformulated with purely real vectors and matrix, as follows:
\begin{equation}
{\bf P}' = {\bf F}'\tilde{\bf P}', \label{eq:obseq2}
\end{equation}
where ${\bf P}' = ({\bf Q},{\bf U})^t$, $\tilde{\bf P}' = (\tilde{\bf Q},\tilde{\bf U})^t$, and the Fourier matrix is reformulated as,
\begin{eqnarray}
{\bf F}' & = &
\left(
\begin{array}{cc}
	{\rm Re}({\bf F}) & - {\rm Im}({\bf F}) \\
	{\rm Im}({\bf F}) &   {\rm Re}({\bf F}) \\
\end{array}     \right),
\end{eqnarray}
where ${\rm Re(z)}$ and ${\rm Im(z)}$ indicate the real and imaginary parts of a complex number $z$.

Equations (\ref{eq:obseq1}) and (\ref{eq:obseq2}) are under-determined linear equations, since the number of the polarization spectrum points $M$ is smaller than the number of FDF pixels $N$. Hence, an infinite number of possible solution exists, even if one can ignore a noise.

\subsection{Regularized Maximum-likelihood Reconstruction}\label{sec:3.2}
Since the observational equation (Equation \ref{eq:obseq2}) is generally ill-posed and with an infinite number of possible solutions, a practical reconstruction is aiming to take the most likely solution based on a reasonable prior assumption matching with properties of the observing sources. Such a process can be mathematically described with a following equation,
\begin{equation}
\min _{\tilde{\bf P}'} f(\tilde{\bf P}')\,\,\,{\rm subject\,\,to}\,\,\,||{\bf P}'  - {\bf F}'\tilde{\bf P}'||_2^2<\varepsilon. \label{eq:regmax1} 
\end{equation}
Here, the term $||{\bf P}'  - {\bf F}'\tilde{\bf P}'||_2^2$ is the traditional chi-square ($\chi ^2({\bf P}'|\tilde{\bf P}')$) term and $\varepsilon$ represents the noise of observations. $f(\tilde{\bf P}')$ is a regularization function, which penalizes (or evaluate) a solution $\tilde{\bf P}'$ based on a given prior assumption for it. Therefore, Equation \ref{eq:regmax1} derives a reasonable solution based on a given prior assumption $f(\tilde{\bf P}')$ within a given residual ($\varepsilon$) between the observational data and solution ($||{\bf P}'  - {\bf F}'\tilde{\bf P}'||_2^2$).

Equation \ref{eq:regmax1} can be transformed to its Lagrangian form,
\begin{equation}
\min_{\tilde{\bf P}'} \left(||{\bf P}'  - {\bf F}'\tilde{\bf P}'||_2^2 + \Lambda  f(\tilde{\bf P}') \right), \label{eq:regmax2} 
\end{equation}
which is minimizing the cost function defined by a sum of the $\chi ^2$-term and the regularization function weighted by a regularization (or hyper) parameter $\Lambda$ connected with the noise $\varepsilon$ in Equation \ref{eq:regmax1}. If one takes a exponential of the cost function, then one can obtain
\begin{equation}
\max_{\tilde{\bf P}'} \left[ \exp \left(-\frac{\chi ^2({\bf P}'|\tilde{\bf P}')}{2}\right)\exp \left( 
-\frac{\Lambda}{2}  f(\tilde{\bf P}')\right) \right], \label{eq:regmax3} 
\end{equation}
which solves a solution maximizing the posterior distribution given by a product of the likelihood (the first factor) and the prior for the solution (the second factor). Therefore, the regularization function can be indeed interpreted as the prior knowledge for the solution in the Bayesian framework. 

For the reconstruction of the FDF, the most important consideration is the choice of a reasonable regularization function, which is fundamentally related to what kind of the prior assumption fits to the observing sources. For radio interferometric imaging, many prior assumptions have been proposed such as the sparsity of the solution (CS-based methods including this work; see Section \ref{sec:3.3}), the information entropy of the solution \citep[Maximum Entropy Methods (MEMs); see][for a review]{narayan1986}, multi-scale patch distributions of the solution \citep{bouman2016}, and a Gaussian-process-based prior \citep{bouman2018}. Very recent work demonstrates that regularized maximum likelihood methods which directly solve Equation \ref{eq:regmax1} or \ref{eq:regmax2} often provide higher fidelity reconstructions than the traditional CLEAN approaches for polarimetric imaging, which shares almost equivalent mathematical formulation with Faraday tomography, by utilizing sparse regularizations \citep{akiyama2017a, birdi2018} or MEM-based functions \citep[e.g.][]{chael2016, coughlan2016}. Since these methods directly solved the observing equations, it can handle multiple types of data \citep[e.g.][]{chael2016, chael2018, bouman2016, bouman2018, akiyama2017a, kuramochi2018} or additional observing effects \citep{johnson2016}.

\subsection{Sparse Reconstruction}\label{sec:3.3}
A simple prior assumption for the regularized maximum likelihood reconstruction is the {\it sparsity} of the solution based on an Occam's razor-like principle that the most simple solution with a minimum number of non-zero components in some basis must be taken. A representation of the sparse reconstruction is given by an optimization problem to minimize the number of nonzero components in the solution $\tilde{\bf P}'$ within a given uncertainty $\varepsilon$ between the observational data $\tilde{\bf P}$ and the solution $\tilde{\bf P}'$. This problem can be mathematically described by
\begin{equation}
\min _{\tilde{\bf P}'} ||\tilde{\bf P}'||_0\,\,\,{\rm subject\,\,to}\,\,\,||{\bf P}'  - {\bf F}'\tilde{\bf P}'||_2^2<\varepsilon. \label{eq:speq} 
\end{equation}
$||x||_p$ is the $\ell _p$ norm of the vector ${\bf x}=\{x_i\}$ given by,
\begin{equation}
||{\bf x}||_{p} = \left( \sum_i |x_i|^p \right) ^{\frac{1}{p}}
\end{equation}
for $p \ge 1$, We defined $||\cdot||_0$ as the number of nonzero components (though this is an abuse of the norm function, it is widely used in CS-related works). 
A direct approach for solving Equation (\ref{eq:speq}) is to optimize a combination, where all possible combinations of zero components of $\tilde{\bf P}'$ will be tried one-by-one. This exhaustive search is extremely computational expensive for the large dimensional $\tilde{\bf P}'$. 

Two major approaches have been taken to overcome this difficulty. One is the greedy approach adopted in the CLEAN (the Matching Pursuit) algorithm and its variants. The greedy approach involves zero-filling process, where $P(\lambda^2)$ is assumed to be 0 for unsampled $\lambda ^2$ points in inverse Fourier transformation from the data to the FDF. Then, sparse solutions are reconstructed by selecting a nonzero element on the FDF domain incrementally and in a greedy manner. CLEAN is often computationally cheap and efficient, although it is often highly affected by the rotation measure spread function (RMSF) caused by zero-filling process.

The other approach is the convex-relaxation adopted in CS-related techniques, which directly solves Equation \ref{eq:speq} with an convex approximation of $\ell_0$-norm. In this approach, $\ell _0$-norm is replaced to $\ell _1$-norm, given by,
\begin{equation}
\min _{\tilde{\bf P}'} ||\tilde{\bf P}'||_1\,\,\,{\rm subject\,\,to}\,\,\,||{\bf P}'  - {\bf F}'\tilde{\bf P}'||_2^2<\varepsilon. \label{eq:speq1} 
\end{equation}
Equation (\ref{eq:speq1}) can be transformed in the Lagrangian form
\begin{equation}
\min _{\tilde{\bf P}'} \left( ||{\bf P}'-{\bf F}'\tilde{\bf P}'||^2_2 + \Lambda _\ell ||\tilde{\bf P}'||_1 \right).
\end{equation}
This is known as the least absolute shrinkage and selection operator \citep[LASSO,][]{tibshirani1996}. 

The LASSO is applied to the Faraday tomography in the pioneering CS work of \citet{li2011b} and also to radio interferometric imaging in \citet{honma2014} and \citet{akiyama2017b}. The convex nature of the $\ell _1$-norm enables to pick up an unique sparse solution from infinite number of possible solutions. $\Lambda _\ell$ adjusts the sparsity of solution, and is related to the observational uncertainty $\varepsilon$. A notable advantage of this approach is that it does not involve zero-filling procedures, and therefore the reconstruction is less affected by the RMSF.

The $\ell _1$-norm of ${\tilde{\bf P}'}$ would be a reasonable regularization function when the sparsity of the FDF is a plausible assumption, for example, for Faraday-thin (i.e. compact) components \citep{li2011b}. On the other hand, if the number of non-zero pixels is not small compared to the number of data points, simple $\ell _1$-norm regularization may reconstruct a too-sparse FDF. This situation can arise when the FDF pixel size is set to be much smaller than the widths of the FDF components (see \citealt{li2011b}; see also \citealt{akiyama2017b} for the case of radio interferometric imaging). One of the promising approaches to 
overcome 
this issue is to change the basis of the image to a more sparse one by replacing $||\tilde{\bf P}'||_1$ to $||{\bf W}\tilde{\bf P}'||_1$, which is convex as well. ${\bf W}$ is a basis transform, for instance, using wavelet transform \citep{li2011b, andrecut2012}. If one knows a reasonable dictionary basis that makes the FDF sparse, then it will give a high-fidelity multi-resolution support of the sparse reconstruction. This approach has been successful for total intensity imaging in the last decade and also very recently presented for polarimetric imaging \citep{birdi2018}. 

The authors have taken another popular approach utilizing additional computationally cheap sparse-gradient regularization providing a piesewise smooth solution. This approach does not require a specific wavelet basis and is based upon an idea that a reasonably smooth (not too sparse) image or FDF may be obtained by adjusting smoothness using regularization of its gradients. A popular convex regularization function for this purpose is the total variation (TV), which is often defined by the gradient of $\ell _ 1$-norm given by
\begin{equation}
||{\bf x}||_{\rm tv} = \sum _j |x_{j+1} - x_j|.
\end{equation}
This provides edge-preserved smooth reconstruction. Meanwhile, astronomical images and/or FDF spectra can be smoothly varied without strong edges. Therefore, \citet{kuramochi2018} proposed to use its variant, total squared variation (TSV) defined by 
\begin{equation}
||{\bf x}||_{\rm tsv} = \sum _j |x_{j+1} - x_j|^2,
\end{equation}
which is also convex and leading to edge-smoothed solutions that more generally fits to astronomical data. We have developed imaging techniques with $\ell _ 1$-norm and TV/TSV, which try to reconstruct a reasonable solution by balancing sparsity of the solution and its gradient \citep{ikeda2016, akiyama2017a, akiyama2017b, kuramochi2018}. In particular, this $\ell _ 1$+TV (or TSV) regularization can be easily extended to full polarimetric imaging \citep{akiyama2017b}, which is mathematically similar to Faraday tomography. In this work, we extend this approach to Faraday tomography as described in Section \ref{sec:3.4}.

Finally, we note a mathematical and computational benefit of sparse reconstruction techniques compared to the MEM-based approaches, which are on another popular branch of prior functions. The important advantage of the sparse reconstruction techniques is the convex nature of the regularized functions even for negative solutions. Popular entropy functions for MEMs are convex only for non-negative solution (e.g., Stokes I) or the absolute of the linear polarization (e.g., $|P|$), and therefore the problem generally becomes non-convex optimization and also requires additional regularization for interferometric polarimetric imaging and Faraday tomography \citep[e.g.][for a review of entropy functions]{chael2016}. 

\subsection{The Proposed Method}\label{sec:3.4}
In this paper, we extend our polarimetric imaging algorithm \citep{akiyama2017b} to Faraday tomography. The proposed method solve the following optimization problem to obtain the FDF,
\begin{equation}
\min _{\tilde{\bf P}'} \left(||{\bf \Sigma}({\bf P}'-{\bf F}'\tilde{\bf P}')||^2_2 +\Lambda_\ell ||\tilde{\bf P}'||_1 + \Lambda_t ||\tilde{\bf P}'||_{\rm tv\,or\,tsv} \right),
\end{equation}
where ${\bf \Sigma}$ is a diagonal matrix given by $\{\Sigma_{ii} = 1/\sigma _ i,\,\,i=1,2,...,2M\}$, representing weights on data due to their uncertainties.

The above problem is convex, and its solution is uniquely determined regardless of initial conditions and regularization parameters. We use a fast iterative shrinkage-thresholding algorithm (FISTA) to solve this problem, which was originally proposed in \citet{beck2009a} for TV regularization and in \citet{beck2009b} for $\ell_1$ regularization. We adopt a monotonic variant of the FISTA described in \citet{akiyama2017b}, which is modified for both $\ell_1$+TV/TSV regularizations.

The most important parameters in the proposed method are the regularization parameters for the $l1$-norm ($\Lambda_{\ell}$) and TV/TSV ($\Lambda_{\rm t}$), which determine the sparseness and smoothness of the FDF, respectively. Too-large regularization parameters provide too-simple images, which do not fit the data well. Meanwhile, too-small regularization parameters lead to fit noises in the data, i.e. overfit the data and give the larger number of non-zero FDF pixels unreasonably, even though it could give better $\chi^{2}$ values.

The determination of the regularization parameters is a common issue in existing techniques for Faraday tomography. For instance, for RM CLEAN, the number of CLEAN components is the issue; the too-large number of CLEAN components would result in the overfit. \citet{miyashita2016} showed that a reduced $\chi ^{2}$, which is conventionally defined as $\chi^{2}/N$, is indeed not always a good indicator for goodness-of-fitting of RM-CLEAN. Therefore, for determining preferable regularization parameters, we need to evaluate some quantities describing goodness-of-fit working as an Occam's razor preventing the overfit.

For evaluating goodness-of-fit, we adopt the cross validation (CV) which works well for the interferometric imaging \citep{akiyama2017a}. Multiple rounds of CV are performed to confront statistical variance, and we choose a 10-fold (ten rounds of) CV. In a 10-fold CV, ten equal-sized subsamples are made by random sampling from the original data. Nine subsamples are used as the training set for performing the model fit, and the remaining single subsample is used as the validation set for testing the model using $\chi^{2}$. We repeat the procedure by changing the subsample for validation data ten times until all subsamples are used for both training and validation. Finally, the ten $\chi^{2}$ values of the validations are averaged and the average is used to determine the optimal regularization parameters which minimize the deviations (e.g. $\chi^{2}$) against the validation sets. As already noted, too-large and too-small regularization parameters give a too-simple and too-complicated models, respectively. Both cases indicate large deviations against the validation set, so that they are safely excluded. The optimal FDF is reconstructed by the full data set using the selected parameter set.

\section{Example Application}\label{sec:4}
\begin{figure*}
\centering
\includegraphics[width=\textwidth]{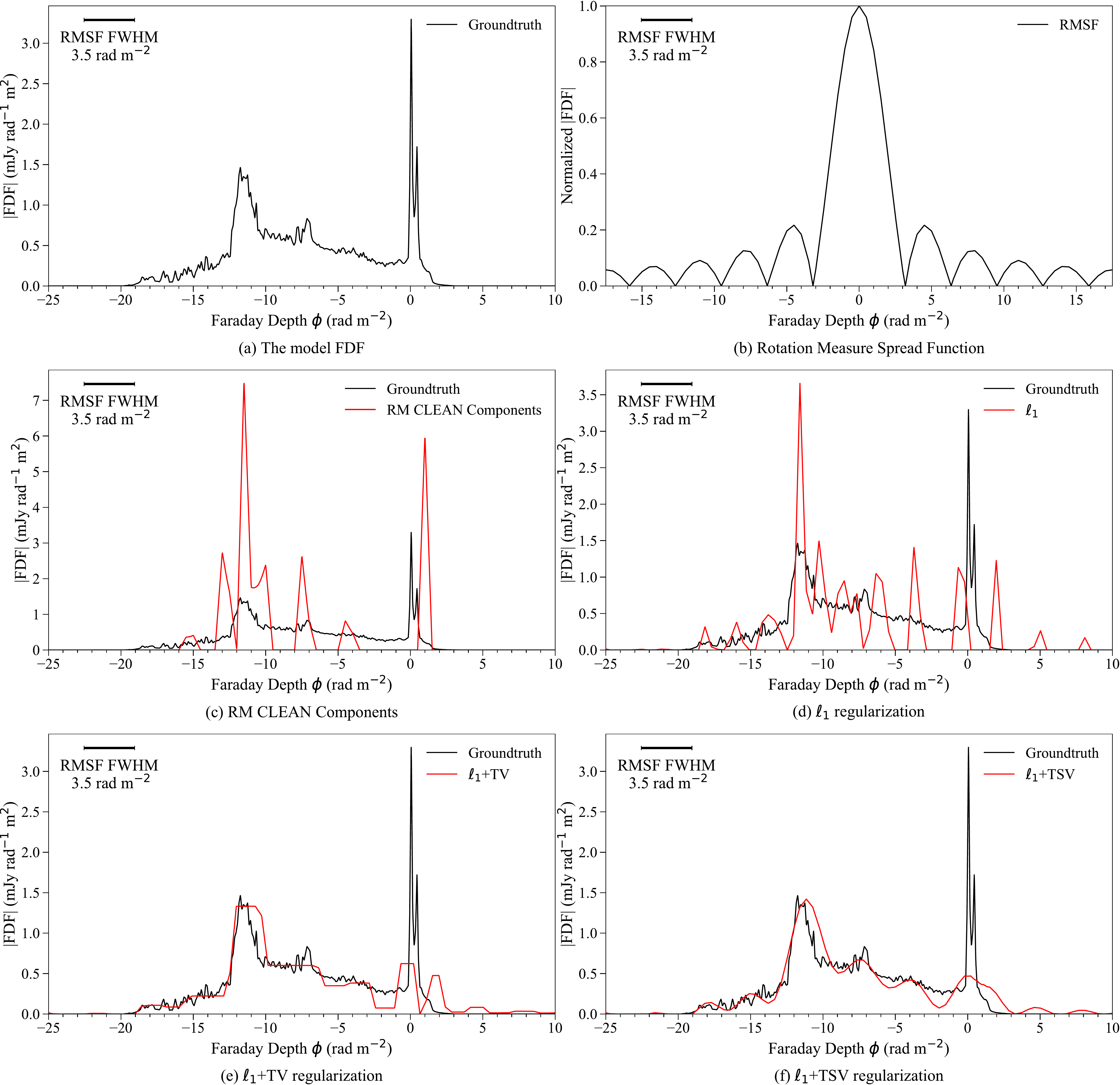}
\caption{Absolute values of FDFs. (a) The model FDF adapted from \citet{ideguchi2014b}. (b) The Rotation Measure Spread Function (RMSF) of the presented synthetic observation. (c-f) The reconstructed FDFs (red lines) compared to the model FDF (the black line) with (c) RM-CLEAN components, (d) $\ell _1$ regularization, (e) $\ell _1$+TV regularization, and (f) $\ell _1$+TSV regularization, respectively. All reconstructions are shown without any post-processed convolution.}
\label{fig:FDFs}
\end{figure*}

As a demonstration of the proposed method, we show an application to a synthetic observation of a galactic FDF model \citet{ideguchi2014b} shown in Figure \ref{fig:FDFs}(a), which is calculated by a small portion of a face-on galaxy using a sophisticated galactic model \citep{akahori2013}. The model has very complicated structures with both Faraday-thick and Faraday-thin components caused by turbulent ISM in a coherent galactic magnetic field, which may be difficult to be approximated by a set of analytic functions like Gaussian.

We made synthetic observational data by Fourier transformation of the FDF model, where the observing frequency of the UHF band, 300 - 3000 MHz, is chosen because of the suitability \citep[see e.g.][]{akahori2018b}. The number of 390 frequency channels are considered, where their spacing is fixed in $\lambda^2$ space\footnote{Although all the techniques adopted in this Section do {\it not} require sample converges equally-spaced in $\lambda^2$ space, we adopted such data to compare more pure performance of reconstructions without any observational bias in the density of frequency samples.}. The resultant Rotation Measure Spread Function (RMSF), the point-spread function of the observation, is shown in Figure \ref{fig:FDFs}(b), which has the FWHM of 3.5~rad~m$^{-2}$. We add random Gaussian noise with the mean of 0 and the standard deviation of 0.1~mJy to the synthetic data at each channel.

\begin{figure*}
\centering
\includegraphics[width=0.6\textwidth]{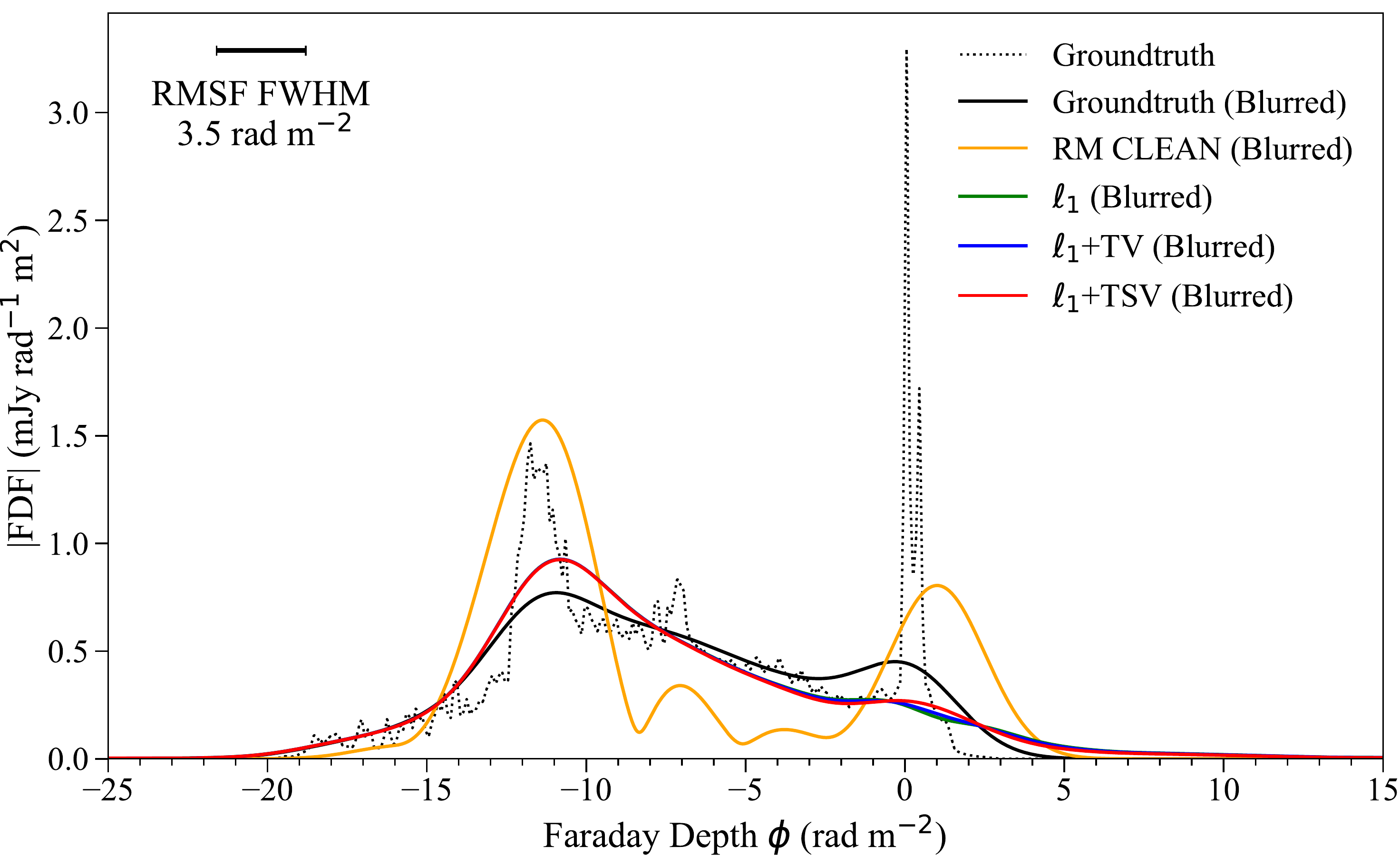}
\caption{The model and reconstructed FDFs blurred with a Gaussian with the FWHM of 3.5~rad~m$^{-2}$ same to the RMSF of the synthetic observation. The model FDF and blurred one are shown in the black dot and solid lines, respectively. The blurred reconstruction is shown in orange, green, blue and red solid lines for RM CLEAN, $\ell_1$, $\ell_1$+TV and $\ell_1$+TSV regularizations, respectively.}
\label{fig:FDFs_blurred}
\end{figure*}

The FDFs are reconstructed from the synthetic data set with multiple methods: RM-CLEAN \citep[e.g.][]{heald2009} with an implementation of \citet{miyashita2016}, $\ell_1$ regularization \citep{li2011b}, $\ell_1$+TV and $\ell_1$+TSV regularization (this work). The Faraday depth range of -100 to 100~rad~m$^{-2}$, larger than the adopted FDF model is adopted with a grid size of 0.50~rad~m$^{-2}$ for RM-CLEAN and 0.43~~rad~m$^{-2}$ for the latter approaches. For the $\ell_1$ and $\ell_1$+TV/TSV regularization, 10-fold CV was adopted to determine the best parameter set from a range of $\Lambda _\ell$, $\Lambda _t$=$10^{-5},\,10^{-4},...,10^5$. The parameter set determined with CV is $\Lambda _\ell=10^{-1}$ for $\ell_1$ regularization, $(\Lambda _\ell ,\Lambda _t) =(10,1)$ for $\ell_1$+TV, $(\Lambda _\ell,\Lambda _t )=(10,10^3)$ for $\ell_1$+TSV regularization. RM-CLEAN is performed with a CLEAN gain of 0.1 with a threshold of 0.1~mJy~rad$^{-1}$~m$^{2}$.

The results are shown in Figure~\ref{fig:FDFs} (c-f). RM-CLEAN and $\ell_1$ regularization, adopting the same prior assumption but with different optimization approaches, reconstructs artificially sparse FDFs. On the other hand, the proposed approach, $\ell_1$+TV and $\ell_1$+TSV regularizations, clearly provide higher-fidelity reconstructions for both the diffuse and peaked components even without post-processing Gaussian convolution often adopted for CLEAN reconstructions \citep{thompson2017}. As expected, the reconstruction with the $\ell_1$+TSV is more edge-smoothed than that with the $\ell_1$+TV. Our approaches resolve structures even less than the resolution in $\phi$ space (super-resolution), as also expected.

In Figure \ref{fig:FDFs_blurred}, we show the model and reconstructed FDFs blurred with a Gaussian whose FWHM is same to that of the RMSF. The blurred FDFs represent those at a nominal Faraday-depth resolution of the synthetic observation, although they are no longer consistent with the observational data. At this resolution, it is clearly shown that RM-CLEAN tends to over estimate two bright components than the blurred groundtruth and poorly reconstruct the diffuse component of the model FDF. On the other hand, CS-based regularized maximum-likelihood approaches provide equally better reconstructions regardless of the regularization functions. This actually demonstrates that the regularized maximum-likelihood approaches are less-affected by the observational Faraday-depth resolution or the RMSF, and provide better reconstructions of complicated structures on scales comparable with the observational resolution.

\begin{figure}
\centering
\includegraphics[width=1.0\columnwidth]{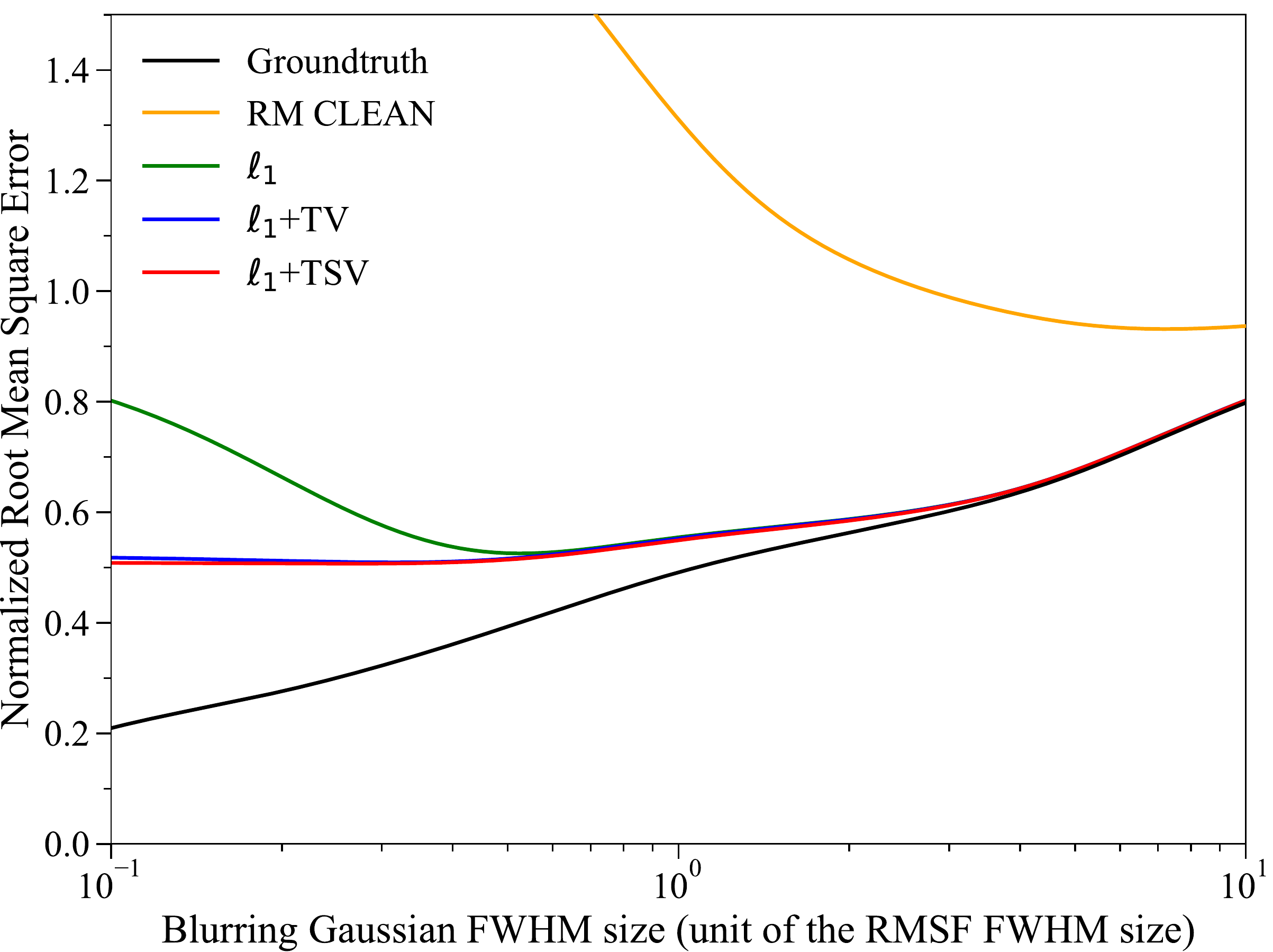}
\caption{The NRMSE between the non-convolved original model FDF and Gaussian-convolved model/reconstructed FDFs, as a function of the FWHM size of the convolving Gaussian beam. The black curve indicates the NRMSE of the model FDF, while other curves indicate the NRMSEs of the reconstructed FDFs. The colors of lines are same to Figure \ref{fig:FDFs_blurred}.}
\label{fig:NRMSE}
\end{figure}

For more quantitative evaluation of the reconstruction performance at each Faraday-depth resolution, we conduct multi-scale error analysis using the normalized root-mean-square error \citep[NRMSE;][]{chael2016} metric, defined by
\begin{equation}
    {\rm NRMSE}(\tilde{{\bf P}},\tilde{{\bf P}}_{\rm ref}) = \sqrt{\frac{\sum_i|\tilde{P}_{i}-\tilde{P}_{{\rm ref},i}|^2}{\sum _i |\tilde{P}_{{\rm ref},i}|^2}},
\end{equation}
where $\tilde{{\bf P}}$ is the full complex FDF to be evaluated and $\tilde{{\bf P}}_{\rm ref}$ is the reference FDF. Here, we adopt the {\it non-convolved} groundtruth (i.e. model) FDF as the reference FDF. Figure \ref{fig:NRMSE} shows the NRMSE metric for the groundtruth and reconstructed FDFs convolved at various Faraday-depth resolutions. The best-case scenario, where the differences from the original FDF is due solely to a loss of resolution and not to errors in reconstruction, is shown by the black curve for the groundtruth FDF. Figure \ref{fig:NRMSE} clearly shows that the CS-based approaches achieve less errors therefore better fidelity over the entire range of spatial scales. 

Both three CS-based approaches show consistent errors from a larger to a smaller scale until about the half of the RMSF FWHM size, then $\ell_1$ regularization shows a rapid increase at smaller scales in the NRMSE metric. This shows that on such small scales the FDF is no longer sparse and its underlying assumption that the FDF can be described with a small number of Faraday-thin components does not work well. On the other hand, $\ell_1$+TV/TSV regularizations show much more modest variations in the superresolution regime. This demonstrates that the prior assumption of these regularization provides better fits to the realistic FDF model adopted in this work. $\ell_1$+TSV provides a slightly better reconstruction than $\ell_1$+TV that tends to create more artificial edge-like features in the reconstructed FDF.

Another important results shown in Figure \ref{fig:NRMSE} are that the reconstructed FDFs with $\ell_1$+TV/TSV provide lower NRMSEs even without any post-processing convolution than those of FDFs convolved at the nominal resolution, while $\ell_1$ and RM-CLEAN do require the post-processing convolution to achieve a better fidelity. Since the proposed method reconstruct smoother FDFs with keeping consistency with observed polarization spectra, the post-processing Gaussian convolution which make the FDF {\it inconsistent} with observational data does not improve the fidelity, and therefore would be no longer required at least for obtaining a higher fidelity for the proposed technique.

\section{Summary and Future prospects}\label{sec:5}
We have presented a new technique for Faraday tomography utilizing $\ell_1$+TV/TSV regularization, which is originally developed for full-polarimetric imaging of radio interferometry. With an example application, we have demonstrated that the proposed technique can indeed achieve a higher performance than the RM-CLEAN approach and previously-proposed pure $\ell_1$ regularization. A multi-scale NRMSE analysis has indicate that the proposed approach provides smaller errors for raw (non-convolved) reconstructed FDFs than reconstructed FDFs convolved at the nominal RMSF FWHM. This indicates that the prior assumptions of the technique is well-matched with the FDF model, and therefore the post-processing Gaussian does not improve the fidelity of the reconstruction. Our results demonstrate that the proposed technique are an attractive choice for Faraday tomography with the next generation broad-band low-frequency radio interferometers. We here describe our future prospects for the proposed technique.

In general, the performance of the FDF reconstruction depends on various factors including the number, the signal-to-noise ratio, and the frequency coverage of the data sample, the complexity of the observing FDF, the range of parameter spaces, and so on. In addition, the performance can be evaluated with more source-specific metrics related to physical quantities that the observers would like to measure. Systematic studies with more source-specific evaluations are necessary to clarify such performance dependence, which are beyond the scope of this paper. The studies will be reported in a series of forthcoming papers (Yamaguchi et al., Miyashita et al., and Ideguchi et al. in preparation).

Although the current approach with simple $\ell_1$ and TV/TSV regularization has already shown a potential to highly improve the fidelity of Faraday tomography with the next generation instruments, its fidelity can be more enhanced with a small correction to the proposed technique. For instance, the current approach regularize each Stokes FDF at $Q$ or $U$ separately, which does not enhance the sparsity better enough to eliminate the components at $\phi >2$ rad~m$^{-2}$ for the presented synthetic observation (see Figure \ref{fig:FDFs}). This would be improved by regularizing $P$ rather than each Stokes parameter. Another promising approach is re-weighted $\ell_1$ regularization recently proposed for polarimetric imaging \citep[polarized SARA;][]{birdi2018}, which has a strong advantage to maximize the sparsity of the reconstruction and effectively eliminate such noise artifacts. In forthcoming work, we will improve the regularization functions and associated optimization approaches.

Another prospect is on accelerating the algorithm for enormous data sets expected to be obtained with the next generation radio interferometers such as the Square Kilometer Array (SKA) and its precursors. This would require acceleration of the proposed algorithm by implementation of (i) a non-uniform fast Fourier transform (NUFFT) and/or a proper frequency-gridding algorithms, (ii) inclusion of major/minor cycles similar to the Cotton-Schwab CLEAN \citep{schwab1984} reducing the number of the non-uniform Fourier transformation, (iii) faster search of the hyper parameters than the current 10-fold CV. For the hyper parameter search, very recently, a heuristic approach to compute validation errors of $N$-fold CV has been developed by \citet{obuchi2016a} for $\ell_1$-regularization and \citet{obuchi2016b} for TV-/TSV-regularization. This approach enables to estimate validation errors for each parameter set only with a single reconstruction using full data sets, which will significantly decrease CV's computational cost. We will implement this algorithm in forthcoming work.

Finally, our paper provides a clear example that recent active developments of the imaging techniques for the next generation interferometers will be also beneficial to the Faraday tomography and mathematically similar linear-inversion problems. Bi-lateral follow-up applications of state-of-the-art techniques intensively developed in both utilities would be continuously important to overcome many challenges for the next generation radio astronomy as well as further improvements of existing techniques.

\vspace{12pt}

\acknowledgements
KA thanks Masato Okada and Mareki Honma for fruitful discussions on applications of sparse modeling to Faraday tomography. KA is a Jansky Fellow of the National Radio Astronomy Observatory. YM is a Research Fellow of the Japan Society for the Promotion of Science (JSPS). The National Radio Astronomy Observatory is a facility of the National Science Foundation operated under cooperative agreement by Associated Universities, Inc. The Black Hole Initiative at Harvard University is financially supported by a grant from the John Templeton Foundation. This work was financially supported in part by a grant from the National Science Foundation (AST-1614868; KA), JSPS KAKENHI Grant Numbers JP25120007 (KA, S.Ikeda), JP24540242, JP25120007, JP25120008, JP15H05896 (KT), JP16H05999 (KT), and JP17H01110 (TA, KT), Bilateral Joint Research Projects of JSPS (KT), and Grant-in-Aid for JSPS Research Fellow 17J06936 (YM).

\end{document}